# Thermally Regenerative Flow Batteries with pH Neutral Electrolytes for Harvesting Low-Grade Heat

Xin Qian[*], Jungwoo Shin, Yaodong Tu, James Han Zhang, and Gang Chen[*]

Department of Mechanical Engineering,

Massachusetts Institute of Technology, Cambridge, MA 02139

Corresponding email:

xinq@mit.edu

gchen2@mit.edu

**Abstract**

Harvesting waste heat with temperatures lower than 100 ºC can improve system efficiency and reduce greenhouse gas emission, yet it has been a longstanding and challenging task. Electrochemical methods for harvesting low-grade heat have attracted research interest in recent years due to the relatively high effective temperature coefficient of the electrolytes (> 1 mV/K) compared with the thermopower of traditional thermoelectric devices. Comparing with other electrochemical devices such as temperature-variation based thermally regenerative electrochemical cycle and temperature-difference based thermogalvanic cells, the thermally regenerative flow battery (TRFB) has the advantages of providing a continuous power output, decoupling the heat source and heat sink and recuperating heat, and compatible with stacking for scaling up. However, TRFB suffers from the issue of stable operation due to the challenge of pH matching between catholyte and anolyte solutions with desirable temperature coefficients. In this work, we demonstrate a pH-neutral TRFB based on $KI/KI_3$ and $K_3Fe(CN)_6/K_4Fe(CN)_6$ as the catholyte and anolyte, respectively, with a cell temperature coefficient of 1.9 mV/K and a power density of 9 µW/cm$^2$. This work also presents a comprehensive model with coupled analysis of mass transfer and reaction kinetics in a porous electrode that can accurately capture the flow rate dependence of power density and energy conversion efficiency. We estimate that the efficiency of the pH-neutral TRFB can reach 11% of the Carnot efficiency at the maximum power output with a temperature difference of 37 K. Via analysis, we identify that the mass transfer overpotential inside the porous electrode and the resistance of the ion exchange membrane are the two major factors limiting the efficiency and power density, pointing to directions for future improvements.

## 1. Introduction

Low-grade heat sources (<100 °C) contain more than half of the global energy rejection,[1] therefore the reuse of low-grade waste heat is a promising route to improve energy efficiency and reduce $CO_2$ emissions. However, due to the distributed nature and the small temperature difference between the heat source and the environment, harvesting low-grade heat has always been a challenging task. In the past few years, electrochemical methods emerged as a promising alternative to thermoelectrics for converting low-grade heat to electricity. This recent pursuit is because the "effective thermopower" (on the order of 1 to 10 mV/K) [2-5] is relatively high compared with thermoelectric materials (~ 200 µV/K).[6] Several different types of devices have been investigated,[7,8] including thermionic capacitors,[9,10] thermogalvanic cells,[11,12] thermally regenerative electrochemical cycles (TREC)[13-15] and thermally regenerative flow batteries (TRFB)[16,17]. Thermionic capacitors rely on the thermodiffusion of ions, also known as the Soret effect. Driven by a temperature gradient applied between the two electrodes, a concentration gradient of cations and anions is created. If the thermal mobilities of cations and anions are mismatched, a net charge profile will build up, which further generates an electric field across the device. Although thermionic capacitors have shown giant Seebeck coefficients, the device cannot operate in a continuous manner. Heat conduction leakage along with the long thermal charging time has led to low efficiency (0.01%).[5,18] Thermogalvanic cells, on the other hand, generate electricity using a single redox pair while the two electrodes are kept at different temperatures. The electrode potential of a general redox reaction $O + ne \leftrightharpoons R$ would depend on temperature, if the entropy change is nonzero. This temperature dependence is quantified by the temperature coefficient $\alpha$,[4] which is related to the entropy change of the cathodic reaction:

$$\alpha = (s_R - s_O)/nF \qquad (1)$$

where $s_R$ and $s_O$ are the partial molar entropies of species $O$ and $R$, and $F$ the Faraday constant. The redox species at each electrode are rebalanced via Soret diffusion (and hence the measured voltage also includes thermodiffusion contributions) through the electrolyte, enabling thermogalvanic cells to operate continuously similar to thermoelectric devices. Yet its efficiency remained very low (~ 0.8% relative to the Carnot efficiency),[19] due to the undesirable heat conduction across the device [20] and the low ionic conductivity compared with electron and holes in thermoelectric materials.[21]

Compared with thermionic capacitors and thermogalvanic cells operating under a temperature gradient, the TREC is more suitable to harvest the temperature difference in time-domain, such as the temperature change between the daytime and the nighttime.[22] A TREC device is similar to a normal battery with cathode and anode materials and can operate in two modes: (1) electrically assisted[13,21] and (2) charging-free.[14] In an electrically assisted device (Figure 1a), the total temperature coefficient of the cell $\alpha_{Cell}$ is the difference between that of the cathode $\alpha_+$ and the anode $\alpha_-$: $\alpha_{Cell} = \alpha_+ - \alpha_-$, and its sign determines the discharge and charge temperature. If $\alpha_{Cell} > 0$, then the TREC battery needs to be discharged at high temperature, and recharged at low temperature (Figure 1a). The charge and discharge temperatures would be swapped for a negative $\alpha_{Cell}$. In charging-free TREC, the standard electrode potentials of cathode and anode redox pairs at room temperature are matched such that the open circuit voltage is close to zero.[14] Either temperature rise or decrease would generate a nonzero electromotive force, driving electrochemical reactions and generating power at both high or low temperatures. The directions of electrochemical reactions are opposite at high and low temperatures so that reactants are regenerated while generating power. Both electrically-assisted and charging-free devices have high relative efficiency to the Carnot limit (10~30 %), depending on the heat recuperating efficiency 50%~85%.[13,14] However, the requirement of periodic temperature changes does not match waste heat sources, which are usually continuous.

Continuously operation of TREC can be achieved via the thermally regenerative flow battery (TRFB) configuration[7,16] device that composes of two flow batteries working at different temperatures, while using pumps to circulate the electrolytes (Figure 1b). In the example of a TRFB with positive temperature coefficient, the flow cell at high temperature works as a galvanic cell providing current and the voltage for the cold flow cell responsible for regenerating the reactants. To ensure continuous operation, the redox active species ideally only exist in the solution, and the electrodes only provide electrons but do not directly participate in redox reactions. In TRFB, one cell serves as a galvanic cell generating current while providing the electromotive force for the other electrolytic cell to regenerate the redox species. Therefore, TRFB does not need any auxiliary power supplies. Another advantage of TRFB is that heat recuperation can be easily achieved with heat exchangers to pre-heat or pre-cool the regenerated electrolytes. For example, Majumdar demonstrated a 15% efficiency relative to the Carnot efficiency can be achieved using TRFB based on $V^{3+}/V^{2+}$ and $Fe(CN)_6^{3-}/Fe(CN)_6^{4-}$.[16]

However, there are several challenges that needs to be resolved before practical implementation of TRFB, which has not received significant attention so far. The most important challenge is the difficulty in simultaneously achieving high value of $\alpha_{Cell}$ while ensuring the stability of electrolytes. Similar to TREC, the cell temperature coefficient of TRFB is calculated as $\alpha_{Cell} = \alpha_+ - \alpha_-$. Ideally, achieving a high absolute value of $\alpha_{Cell}$ requires not only large absolute values of the temperature coefficients, but also opposite signs of temperature coefficients of the catholyte and the anolyte.

We can understand the routes to high $\alpha_{Cell}$ qualitatively using Born's solvation model. According to Born's solvation model,[23,24] the entropy change of the reaction $O + ne \leftrightharpoons R$ is proportional to the change of squared valence number $z^2$ divided by the ionic radii $r$:

$$\alpha = \frac{s_R - s_O}{nF} \propto \left(\frac{z_O^2}{r_O} - \frac{z_R^2}{r_R}\right) \tag{2}$$

where $s$ denotes the partial molar entropy of the oxidized species $O$ and the reduced species $R$, and $F$ is the Faraday constant. In many redox reactions like $Fe^{3+}+e \leftrightharpoons Fe^{2+}$, the change of the ionic radii is not large enough to affect the sign of $\alpha$. For cation redox pairs, the reduced ion species usually has smaller valence value, $z_O^2 - z_R^2 > 0$, giving rise to a positive $\alpha$. In contrast, anion redox pairs usually have a negative $\alpha$ according to Born's solvation model. Born's solvation model also indicates that higher valence ions with smaller radii tend to have relatively large temperature coefficients. However, another factor that needs to be simultaneously considered is the hydrolysis of these redox ions, which affects the stability of the electrolyte solution. For example, metallic cations like $V^{3+}/V^{2+}$ and $Fe^{3+}/Fe^{2+}$ are good candidates for relatively large positive temperature coefficients,[16] but these solutions are only stable in low pH environments.[25,26] Anions like $Fe(CN)_6^{3-}/Fe(CN)_6^{4-}$, however, are usually stable in neutral or basic environments. Such a mismatched stable range of pH greatly affects the stability of TRFB cells. For example, although pairing cations such as $Fe^{3+}/Fe^{2+}$ and $V^{3+}/V^{2+}$ with $Fe(CN)_6^{3-}/Fe(CN)_6^{4-}$ can achieve a high value of $|\alpha_{Cell}| \sim 2.7$ mV/K and 3.1 mV/K,[16] the proton exchange across the ion exchange membrane would gradually shift the pH value on the $Fe(CN)_6^{3-}/Fe(CN)_6^{4-}$ side, resulting in a complicated decomposition of the $Fe(CN)_6^{3-}/Fe(CN)_6^{4-}$ ions releasing toxic $CN^-$ once the solution becomes acidic.[27] Such cells can only operate stably until the pH buffer is consumed. In addition to the pH matching problem, it is also preferred that the redox active species have the same sign of charges.

For example, the TRFB based on $Cu(NH_3)_4^{2+}/ Cu(NH_3)_2^+$ and $Fe(CN)_6^{3-}/Fe(CN)_6^{4-}$ has a high temperature coefficient $|\alpha_{Cell}| = 2.9$ mV/K;[17] however, the different sign of charges between the reactive ion pairs makes it difficult to find the optimal ion exchange membrane to prevent the cross-over of redox active ions.[17]

In addition to the challenges of pH matching and electrolyte stability, there also lacks a comprehensive analysis of coupled mass transfer and electrochemical kinetics. For flow batteries, the steady-state current density is determined by the rate of redox reaction at the electrode surface, which is proportional to the local concentration of redox species at the liquid-electrode interfaces.[28] When the mass transfer is slow compared with the redox reaction rate, there could be a large concentration gradient near the electrode, resulting in concentration overpotential and limiting the maximum current of the cell. When reactants are depleted locally at electrode surfaces, the current reaches the limiting current. Clearly, such mass transfer process and thus the discharging current depends on flow rates. Another factor that can limit the reaction rate is the polarization of electrodes, which becomes important when a porous electrode is used in the battery. Consider a cathodic reaction with $n$ electrons transferred from the electrode to the electrolyte. Charge conservation dictates that this amount of extra negative charges needs to be compensated by the ionic current (Figure 1b). If the ionic diffusion in the porous electrode is slow, the total current that can be extracted from TRFB would also be limited, which is known as electrode polarization. All these kinetic processes described would result in an additional potential drop at the electrode/electrolyte interface, known as the overpotential. However, the analysis of TRFB in the literature either neglected the overpotential[29] or assumed an infinitely large mass transfer coefficient.[16] In addition, these irreversible losses are usually coupled with the forced convection of electrolytes, dictating the flow-rate dependent power density. It is therefore necessary to comprehensively study the coupled effects of mass transfer and reaction kinetics at different flow rates for optimized performance.

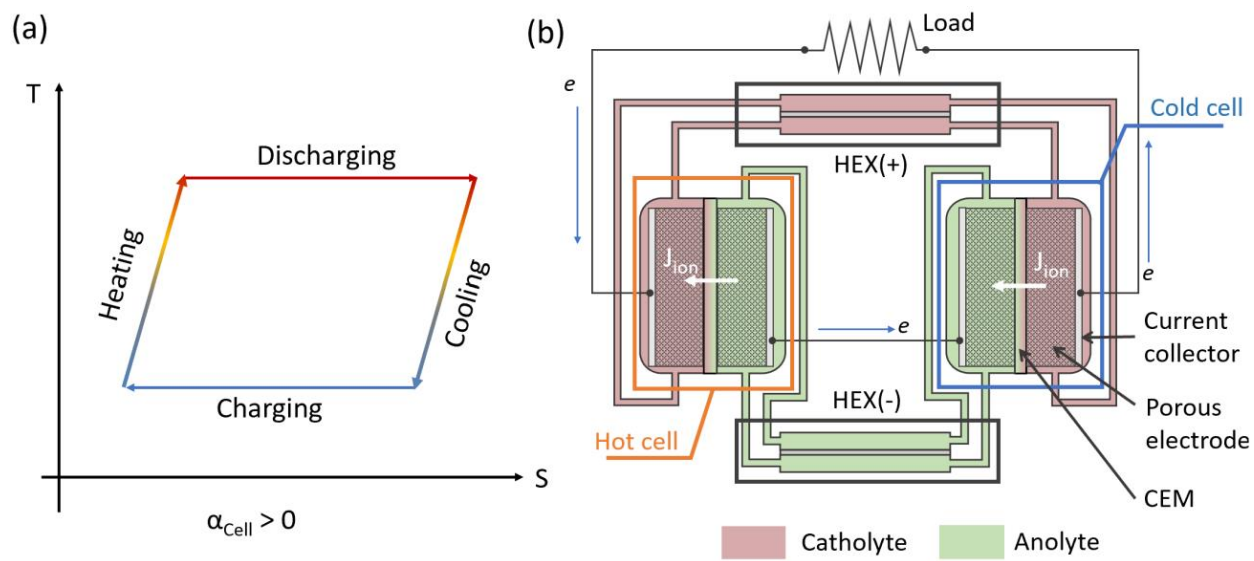

Figure 1. (a) Temperature-entropy diagram of the thermally regenerative electrochemical cycle (TREC) for positive temperature coefficient $\alpha_{Cell} > 0$. If the temperature coefficient of the cell is negative, then the charging takes place at the higher temperature while the discharging is performed at lower temperature. For charging-free TREC, both high temperature and low temperature reactions can output power, but directions of electrochemical reactions are opposite. (b) Schematic of thermally regenerative flow battery (TRFB) with $\alpha_{Cell} > 0$. Catholyte (or anolyte) is the electrolyte that is reduced (or oxidized) in the discharging cell, which is the cell operating at higher temperature for $\alpha_{Cell} > 0$. The flow direction of electrons is indicated by the blue arrow, while the ionic current ($J_{ion}$) across the cation exchange membrane (CEM) is indicated by the white arrows. Two heat exchangers (HEXs) are implemented for heat recuperation of the catholyte (+) and the anolyte (-).

Here we tackle the challenges aforementioned and demonstrated a pH neutral TRFB using $KI/KI_3$ as the catholyte and $K_3Fe(CN)_6/K_4Fe(CN)_6$ as the anolyte with all-anionic redox active ion pairs. The redox reaction of catholyte between the iodide ion ($I^-$) and triiodide ion ($I_3^-$) is a two-electron process:

$$I_3^- + 2e \rightleftharpoons 3I^-, \quad E^0 = 0.53 \text{ V} \tag{3}$$

while the redox reaction for the anolyte is written as:

$$Fe(CN)_6^{3-} + e \rightleftharpoons Fe(CN)_6^{4-}, \quad E^0 = 0.37 \text{ V} \tag{4}$$

where $E^0$ denotes the standard electrode potential with respect to the standard hydrogen electrode (SHE). Such pairing of electrolytes only involve anions as the redox active species while sharing the same $K^+$ counterion, allowing convenient cation exchange without shifting the pH value and preventing the proton induced decomposition of $K_3Fe(CN)_6/K_4Fe(CN)_6$. This pH neutral TRFB has a temperature coefficient of 1.9 mV/K and a maximum power density of 9 µW/cm² at the high limit of flow rates. To analyze the energy conversion efficiency and factors affecting the performance of TRFB, we develop a comprehensive porous electrode model for TRFBs, which can capture the effect of flow-rate, mass transfer, surface kinetics, and electrode polarization on the power density and efficiency. We estimate that the maximum efficiency can reach ~ 11% relative to the Carnot limit at a flow rate of 1 µL/min, assuming 90% of heat recuperation. We also pinpoint the mass transfer inside the porous electrode, membrane resistance and heat recuperation effectiveness as the key factors for improving efficiency and power factors in the future.

## 2. Theoretical Model of TRFB with Porous Electrode

Figure 2 shows the geometry of the porous electrode considered in the model. For the convenience of notation, we consider a general half-cell reaction:

$$\nu_O O + ne \leftrightharpoons \nu_R R \tag{5}$$

where $O$ denotes the oxidized species and $R$ denotes the reduced species, with $n$ electrons transferred per unit reaction, $\nu_O$ and $\nu_R$ the stochiometric number. The electrochemical potential of the electrolyte at thermal equilibrium is determined by the concentration of reactants through the Nernst equation:

$$E_{eq} = E^{0\prime} + \frac{\mathcal{R}T}{nF} \ln \frac{c_O^{\nu_O}}{c_O^{\nu_R}} \tag{6}$$

where $\mathcal{R}$ is the universal ideal gas constant, the $c_k = C_k/C^0$ ($k = O, R$) denotes the dimensionless concentration of species to the standard concentration $C^0$ (1 M), and $E^{0\prime} = E^0 + \frac{RT}{nF} \ln \frac{\gamma_O^{\nu_O}}{\gamma_R^{\nu_R}}$ is the formal potential with $\gamma$ the activity coefficients and $E^0$ the standard electrode potential. Apparently, the electrochemical potential of the electrode $\phi_1$ would equilibrate with that of the electrolyte $\phi_2$ when the discharging current is zero: $\phi_1 = \phi_2 = E_{eq}$. However, when the discharging current is nonzero, the equilibrium condition is broken resulting in a kinetic potential drop across the interface, known as the overpotential ($\eta = \phi_1 - \phi_2$). In addition, the equilibrium potential becomes location dependent because the concentration profile inside the bulk is nonuniform at a finite discharging current. Such a nonuniform concentration profile further affects the surface kinetics and mass transfer, resulting in a nonuniform profile of overpotential.

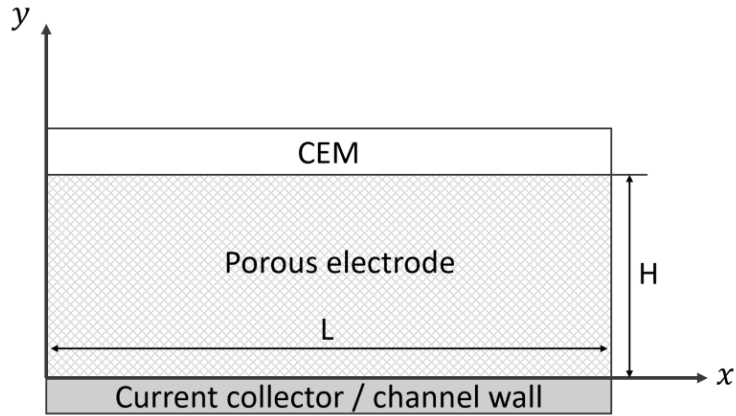

Figure 2. Schematic of the porous electrode sandwiched between the CEM and the current collector or the channel wall. The flow direction is along the $x$-axis.

This section derives a model to determine the overpotential drop inside the porous media as a combined result of mass transfer, electrode polarization and surface reaction kinetics. Section 2.1 outlines the mass transfer inside the porous electrode. An empirical mass-transfer relation for interdigitated media is used to consider the local concentration gradient between the mean concentration and the microscopic concentration at interface between electrolyte and the fibers of the electrode. In Section 2.2, the extended Butler-Volmer equation and condition of charge neutrality are used as the bridges coupling the effects of local mass transfer, surface reaction kinetics and electrode polarization. In Section 2.3, we derive the self-consistent equation for determining the discharge current and outlined the evaluation of efficiency.

**2.1 Mass Transfer in the porous electrode.**

We start deriving the TRFB model by considering spatial profile of concentration, which can be obtained by solving the continuity equation:

$$\frac{d}{dx}\left(-\widetilde{D}_k \frac{dC_k}{dx} + u_s C_k\right) = \pm \nu_k R_V \tag{7}$$

where $R_V$ is the rate of reaction that takes the positive sign for the cathodic reaction per unit volume. Eq. (7) can be obtained by integrating the two-dimensional equation of mass conservation along the $y$-direction, with the boundary conditions that $y$-component of the ionic diffusion currents at $y = 0$ and $y = L$ are zero, because active species cannot penetrate across the current collector or membrane. Such treatment is reasonable because the convective transfer along the $x$-direction is dominant over the diffusion process along $y$, with the dimensionless Peclet number $\text{Pe}_\text{H} = u_s H/\widetilde{D}$ estimated on the order of $10^3$. The subscript $k$ ($= O$ or $R$) denotes the redox species, $\nu$ the stochiometric number and $u_s$ the superficial velocity of the electrolyte. $\widetilde{D}_k$, the effective diffusivity inside the porous electrode, can be related to the diffusivity of the bulk solution $D_k$ through $\widetilde{D}_k = \epsilon D_k/\mathcal{T}$, where $\epsilon$ is the porosity and $\mathcal{T}$ the tortuosity factor of the porous media. The sign of the source term takes $+$ when $k = O$ and the $-$ when $k = R$. By nondimensionalizing the concentration $c_k = C_k/C^0$ and the coordinate $X = x/L$ with $L$ the electrode length, Eq. (7) can be written as:

$$\frac{d^2 c_k}{dX^2} - \text{Pe}_k \frac{dc_k}{dX} + \dot{g}_k = 0 \tag{8}$$

where $\text{Pe}_k = u_s L/\widetilde{D}_k$ is the Peclet number of species $k$ and $\dot{g}_k$ is the normalized rate of generating the species $k$ inside the porous electrode:

$$\dot{g}_O = -\nu_O \frac{R_V L^2}{\widetilde{D}_O C^0}, \qquad \dot{g}_R = \nu_R \frac{R_V L^2}{\widetilde{D}_R C^0} \tag{9}$$

The concentration profile can be solved analytically with the boundary condition $c_k(X = 0) = c_k^0$ and $\frac{d}{dX} c_k(X = 1) = 0$:

$$c_k(X) = c_k^0 + \frac{\dot{g}_k}{\text{Pe}_k}X - \frac{\dot{g}_k}{(\text{Pe}_k)^2}\left[\exp(-\text{Pe}_k(1-X)) - \exp(-\text{Pe}_k)\right] \approx c_k^0 + \frac{\dot{g}_k}{\text{Pe}_k}X \qquad (10)$$

In TREC flow batteries, the superficial velocity of electrolyte is on the order of 0.1~1 mm/s, yielding a large Peclet number $10^3$~$10^4$, therefore the mean concentration profile is approximately linear along the flow direction.

At steady-state, the rate of consuming and generating reactants should also be balanced with the flux of reactants from the bulk solution to the electrode surface, therefore:

$$v_O R_V/A_e = \mathcal{M}_O(C_O - C_O^s), \qquad v_R R_V/A_e = \mathcal{M}_R(C_R^s - C_R) \qquad (11)$$

where $\mathcal{M}_O$ and $\mathcal{M}_R$ are the mass transfer coefficient, and $C_{O,R}^s$ is the mean concentration at the surface of the electrode. $A_e$ is the specific electrode area per unit volume, determined by the porosity $\epsilon$ and fiber diameter $d_f$ as $A_e = 4(1-\epsilon)/d_f$.[30] The dimensionless form of Eq. (10) can be written as:

$$-\zeta \dot{g}_O = \text{Sh}_O(c_O - c_O^s), \qquad \zeta \dot{g}_R = \text{Sh}_R(c_R - c_R^s) \qquad (12)$$

where $\text{Sh} = \mathcal{M} d_H/D$ is the Sherwood number of mass transfer, $d_H$ is the hydrodynamic diameter for the porous media, calculated as $d_H = 0.471\frac{\epsilon}{1-\epsilon}d_f$, and $\zeta = d_H/A_e L^2$ is a geometric factor of the porous electrode.[31] The Sherwood number of the interdigitated flow field inside the porous electrode can be described by:[32]

$$\text{Sh} = 0.018\text{Re}^{0.68}\text{Sc}^{0.50} \qquad (13)$$

where $\text{Re} = \rho u_s d_H/\epsilon\mu$ is the Reynolds number and the $\text{Sc} = \mu/\rho D$ is the Schmidt number where $\mu$ is the dynamical viscosity.

## 2.2 Overpotential and Current Density.

The extended Butler-Volmer equation is the key to correlate the current density to the overpotential loss, surface concentration and the local current density ($J$):

$$J = J_0 \left[ \frac{C_O^S}{C_O} \exp\left( \alpha_S \frac{nF}{RT} \eta \right) - \frac{C_R^S}{C_R} \exp\left( -(1-\alpha_S) \frac{nF}{RT} \eta \right) \right] \tag{14}$$

where $\alpha_S$ is the symmetry factor of the excitation energy barrier, and $J_0 = nFk_0 C^0 c_O^{1-\nu_O \alpha_S} c_R^{\nu_R \alpha_S}$ is the exchange current density with $k_0(T)$ the rate constant calculated through the Arrhenius law:

$$k_0(T) = k_0(T^0) \exp\left[ -\frac{\Delta G}{R} \left( \frac{1}{T} - \frac{1}{T^0} \right) \right] \tag{15}$$

Eq. (14) converges to the original Butler-Volmer equation when $C_O^S = C_O$ and $C_R^S = C_R$:

$$J \xrightarrow{Sh \to \infty} J_0 \left[ \exp\left( \alpha_S \frac{nF}{RT} \eta \right) - \exp\left( -(1-\alpha_S) \frac{nF}{RT} \eta \right) \right] \tag{16}$$

which is valid when the mass transfer coefficients approach infinity (Sh → ∞).

The current density $J$ is evaluated per unit electrode surface area, and can be related to the volumetric reaction rate through charge balance across the electrode/electrolyte interface:

$$nFR_V = JA_e \tag{17}$$

The charge balance condition in the electrolyte should also be considered along the normal direction to the membrane ($y$-axis). For each mole of cathodic reaction happening, the amount of negative charges transferred to the electrolyte is $nFR_V$ per unit volume, which needs to be balanced with ionic current:

$$nFR_V = \nabla \cdot \boldsymbol{J}_{ion} \approx \sigma_{eff} \frac{d^2 \eta}{dy^2} \tag{18}$$

with the divergence of ionic current written as $\nabla \cdot \boldsymbol{J}_{ion} = -\sigma_{eff}\nabla^2\phi_2 = \sigma_{eff}\nabla^2\eta$, using the Ohm's law. For cells with flow-by geometry, the $x$-derivatives of the Laplacian $\nabla^2$ can be neglected since the length of the electrodes is orders of magnitude larger than the thickness, therefore the last approximating equality in Eq. (18) holds. The potential variation inside the conductive electrode ($\phi_1$) is neglected. $\sigma_{eff}$ is the effective ionic conductivity of the electrolyte inside the porous media, estimated as $\sigma_{eff} = \sum_j \frac{F^2}{RT} z_j^2 \widetilde{D}_j C_j$. Together with Eqs. (11), (14), (17) and (18), we can obtain an ordinary differential equation (ODE) considering the electrode polarization:

$$\frac{d^2\eta}{dy^2} = \frac{\exp\left(\alpha_S \frac{nF}{RT}\eta\right) - \exp\left(-(1-\alpha_S)\frac{nF}{RT}\eta\right)}{\frac{\sigma_{eff}}{A_e}\left(\frac{1}{J_0} + \frac{\nu_O}{\mathcal{M}_O C_O nF}\exp\left(\alpha_S \frac{nF}{RT}\eta\right) + \frac{\nu_R}{\mathcal{M}_R C_R nF}\exp\left(-(1-\alpha_S)\frac{nF}{RT}\eta\right)\right)} \quad (19)$$

Using the nondimensionalized coordinate $Y = y/H$ with $H$ the thickness of the porous electrode, and the nondimensionalized overpotential $\psi = \frac{nF}{RT}\eta$, we can derive in the following equivalent ODE:

$$\frac{d^2\psi}{dY^2} = \frac{\exp(\alpha_S n\psi) - \exp(-(1-\alpha_S)n\psi)}{\mathcal{K}\left[\frac{1}{J_0} + \zeta\left(\frac{\nu_O \mathrm{Da}_O}{\mathrm{Sh}_O c_O}\exp(\alpha_S n\psi) + \frac{\nu_R \mathrm{Da}_R}{\mathrm{Sh}_R c_R}\exp(-(1-\alpha_S)n\psi)\right)\right]} \quad (20)$$

where $j_0 = c_O^{1-\nu_O \alpha_S} c_R^{\nu_R \alpha_S} = J_0/J_0^0$ is the dimensionless exchange current, with $J_0^0 = nFk_0 C^0$. $\mathrm{Da}_k$ is the Damköhler number characterizing the relative ratio between the reaction rate to the mass diffusion rate, defined as $\mathrm{Da}_k = k_0 A_e L^2/\widetilde{D}_k$. $\mathcal{K}$ is the dimensionless conductivity $\mathcal{K} = \sigma_{eff} \cdot (RT/F)/(J_0^0 A_e H^2)$, which can be interpreted as the ratio of conduction current driven by the "thermal voltage" $RT/F$ to the exchange current $J_0^0$ at standard concentration. The dimensionless ODE (Eq. (20)) clearly shows that the overpotential is a combined result of mass transfer quantified by the Sherwood number Sh and reaction kinetics characterized by the Damköhler number Da, and the activities of redox species $c_O$ and $c_R$.

Finally, we consider the boundary conditions for solving the overpotential. At $y = H$, the ionic current across the membrane should equal to the discharging current of the cell, and at $y = 0$ where the electrode is in contact with either the current collector or wall of flow channel, the ionic current should be zero. The boundary condition for solving the Eq. (18) or (19) can therefore be written as:

$$\sigma_{eff}\left(\frac{d\eta}{dy}\right)_{y=0} = 0, \quad \sigma_{eff}\left(\frac{d\eta}{dy}\right)_{y=H} = i_{cell}/A \tag{21}$$

where $i_{cell}$ is the discharging current of TRFB and apparent $A$ is the electrode area.

For each location $x$ along the flow direction, we solve the ODE (Eq. 19) using the boundary value problem solver bvp4c of Matlab to obtain the overpotential profile $\eta(x, y)$. In this work, we assume the electrode is conductive enough and neglected the electric potential variation in the solid phase, hence the overpotential drop inside the electrode is then taken as the mean value across the entire electrode domain.[33] Similarly, the electrochemical potential $E(x)$ of the electrolyte is calculated with the local concentrations $C_O(x)$ and $C_R(x)$, and we take the electrode potential as the mean value of the $E(x)$.[33] Rigorously, a full set of two-dimensional hydrodynamic, mass transfer and polarization model should be solved, but we shall see later such the mean-field approach taken in our modeling can already achieve reasonable agreement with experiments and capture the coupled effects of mass transfer, electrochemical kinetics and electrode polarization.

**2.3 Self-Consistent Solution of Discharging Current.**

After discussing the mass transfer and the overpotential inside the electrolyte, we now solve the total current that can be extracted from the cell. We first consider the discharging cell operating at $T_1$, with the electrode potential of cathode (+) and anode (-) expressed as:

$$E_\pm(T_1) = E(T^0) + \alpha_\pm(T_1 - T^0) \tag{22}$$

For convenience, we denote the solution to the ODE of overpotential (Eq. 17) as $\eta = f(R_V)$, then the overpotential at the two electrodes for the discharging cell are expressed as

$$\eta_1^\pm = f(\pm \frac{i_{cell}}{n_\pm F \mathcal{V}}) \tag{23}$$

where the subscript 1 indicates the discharging cell operating at $T_1$, $\mathcal{V}$ the volume of the porous electrode, $n_\pm$ the electron transfer number for the redox reaction of the catholyte (+) and anolyte (-) correspondingly. The anode will adopt a negative sign when computing the volumetric reaction rate since we define cathodic reaction as the positive direction. At a finite discharging current, the voltage for the discharging cell ($V_1$) is therefore

$$V_1(i_{cell}) = V_{oc}(T^0) + \alpha_{Cell}(T_1 - T^0) - (\eta_1^+ - \eta_1^-) - i_{cell} \cdot R_{in} \tag{24}$$

where $V_{oc}(T^0) = E_+(T^0) - E_-(T^0)$ is the open circuit voltage at reference temperature, and $R_{in} = R_{mem} + R_\Omega$ is the internal resistance as a combination of the membrane resistance $R_{mem}$ and electrical Ohmic resistance $R_\Omega$ of the electrode and current collector. Note here we did not lump the overpotential into the internal resistance, since the overpotential loss depends on discharging current and cannot be simply understood as a resistive component. From the extended Butler-Volmer equation, the sign of the overpotential is the same as the sign of current and $R_V$, thereby $\eta_1^- < 0$, and the total overpotential for the cell is positive $(\eta_1^+ - \eta_1^-) > 0$.

The voltage of the charging cell operating at $T_2$ can be computed in the same manner by simply replacing $i_{cell} \to -i_{cell}$, since all the chemical reactions are opposite to the discharging cell. The overpotential for the cathode and the anode can be written as:

$$\eta_2^\pm(i_{cell}) = f(\mp \frac{i_{cell}}{n_\pm F \mathcal{V}}) \tag{25}$$

The voltage of the charging cell at finite current is therefore:

$$V_2(i_{cell}) = V_{oc}(T^0) + \alpha_{cell}(T_2 - T^0) - (\eta_2^+ - \eta_2^-) + i_{cell} R_{in} \tag{26}$$

Finally, the measured potential for the full TRFB is:

$$V(i_{cell}) = V_1 - V_2 = \alpha_{Cell}(T_1 - T_2) - \eta_{tot}(i_{cell}) - 2 i_{cell} R_{in} \tag{27}$$

where $\eta_{tot}(i_{cell}) = (\eta_1^+ - \eta_1^-) - (\eta_2^+ - \eta_2^-)$ is the total overpotential drop. Similarly, $(\eta_2^+ - \eta_2^-) < 0$ and hence the total overpotential drop is positive. In addition, the current is also determined by the load $R_{load}$ in the external circuit, hence the current can be determined by the following equation:

$$\frac{\alpha_{Cell}(T_1 - T_2) - \eta_{tot}(i_{cell})}{(2R_{in} + R_{load})} = i_{cell} \tag{28}$$

Eq. (28) can be solved self-consistently using a minimization solver of Matlab. Finally, the efficiency $\eta_E$ of the TRFB system can be determined as:[34]

$$\eta_E = \frac{V(i_{cell}) \cdot i_{cell} - P_{pump}}{\alpha_{Cell} \cdot \max[T_1, T_2] + \sum_{j=\pm}\left[(1 - \epsilon_{HX,j}) \cdot \rho_j c_{p,j} Q_j (T_1 - T_2)\right] - i_{cell}^2 R_{in}} \tag{29}$$

where the subscript $j$ in the denominator sums over catholyte and anolyte, $\epsilon_{HX}$ the effectiveness of heat recuperation of, $\rho$ and $c_p$ the density and specific heat of the electrolytes, and $Q$ the flow rate. The pump work $P_{pump} = \Delta p \cdot (Q_+ + Q_-)$, where $\Delta p$ is pressure drop along the flow direction inside the porous media is estimated using the Ergun equation: [35]

$$\Delta p = \frac{150\mu L}{d_f}\frac{(1-\epsilon)^2}{\epsilon^3}u_s + \frac{1.75\rho L}{d_f}\frac{1-\epsilon}{\epsilon^3}u_s^2 \tag{30}$$

where $\mu$ is the dynamic viscosity of the electrolytes. In our work, the flow rate across the cell is small and the pump work is negligible compared with the output power of the TRFB.

The porous electrode model of TRFB presented here will be used to comprehensively analyze the factors and sources limiting the power density and efficiency in the following sections, including operating factors such as flow rates, heat recuperation effectiveness and temperature differences, as well as the electrolyte and electrode properties such as rate constants and ionic conductivities.

### 3. Experimental Implementation of the TRFB

In this section, we describe the experimental details of the TRFB testing system. As shown in Figure *3*(a-b), the testing system was composed of two flow batteries, two heat exchangers and a heating box for temperature control. As shown in Figure *3*b, the heating box chamber was built with pink insulation foam and filled with extra ceramic fiber insulation blankets at the top, the bottom and the backside walls. Heating was performed with a coil heater and a heating plate placed at the bottom of the box. The heating plate consisted of an OMEGA silicone fiberglass heater attached to a piece of aluminum plate. The coil heater and the heating plate was connected in parallel to a programmable power supply (BK precision 9130). A thermocouple was suspended in the chamber to monitor the temperature. Temperature control was achieved via a home-built PID model through LABVIEW.

Figure *3*(c-d) shows the assembly of the flow cell, with the flow channels created by cutting the silicone gaskets (2.5 mm thick) filled with the graphite felt (AvCarb® G200, with fiber diameter ~ 10μm) [36] as the porous electrode. The graphite felt was cut into dimensions of 1.3 cm

× 6 cm × 2.5 mm to fit into the flow channel. To improve the hydrophilicity, the graphite felt was heated up to 400 °C in ambient condition up to 24 hours. A Nafion 115 cation exchange membrane was used as the separator between the catholyte and the anolyte to prevent cross over of negatively charged redox species and short circuiting. Each porous electrode was in contact with a strip of Titanium foil (grade II, 25 μm) as the current collector. A thermocouple coated with epoxy was embedded in the flow channel of the anolyte $Fe(CN)_6^{3-}/Fe(CN)_6^{3-}$ to monitor the temperature of the cells. We did not embed the thermocouple into the channel of $I_3/I^-$ solutions, because the dissolved iodine could penetrate through the thin layer of epoxy coating and corrode the thermocouple junction. Gaskets, electrodes and membrane were compressed with two pieces of acrylic plates. Heat exchangers were assembled in a similar way as shown in Figures 3(e-f). At relatively high flow rates (340 μL/min), the heat exchanger had a reasonable effectiveness (~ 90%), estimated by the temperature readings of the four thermocouples embedded inside the flow channels of the heat exchangers. However, due to the small heat capacitance rate, the heat loss across the tube walls was significant, especially at smaller flow rates. The temperature difference between the inlet temperature of hot channel and the hot cell could reach ~20 °C when the hot cell temperature was higher than ~55 °C, despite the insulation foam surrounding the heat exchangers. However, this problem of insulating the tubes would not be as significant when flow rates are scaled up by stacking multiple flow cells. Therefore, we assumed 90% of the heat recuperation effectiveness when estimating the efficiency of the TRFB devices, which is also the typical effectiveness for heat exchangers for liquid-liquid heat transfer. A peristaltic pump was used to inject the electrolytes into the flow cells. The anolyte solution contained 0.375M $K_3Fe(CN)_6$ and 0.375M $K_4Fe(CN)_6$ and the catholyte solution contained 0.1 M $KI_3$ an 1 M KI. The catholyte solution was prepared by dissolving 0.1 M iodine ($I_2$) into 1.1 M of KI solution. Both the catholyte and anolyte chosed are stable near neutral pH and can be paired without the pH buffer solutions. In addition, the active ions of both solutions are anionic, therefore can be well separated using a cation exchange membrane.

We used a potentiostat (BioLogic VSP-300) to characterize the cell temperature coefficient $\alpha_{Cell}$, open circuit voltage and the discharging performance. The current collectors of the anodes (Fe(CN)$_6^{3-}$/ Fe(CN)$_6^{3-}$) of the two cells were connected by a wire, and the current collector of the cathode (I$_3^-$/ I$^-$) of the hot cell was connected with the working electrode terminal of the potentiostat, and the other cathode of the cold cell served as the counter and reference electrode. Before heating, we maintained the system at room temperature for nearly 20 minutes for equilibration of temperature and the relaxation of the initial built in voltage. Then we increased the temperature in a stepwise manner and recorded the open circuit voltage as shown in Figure 4a. By linear regression, the cell temperature coefficient was extracted as $\alpha_{cell} = 1.9$ mV/K (Figure 4b), with only slight deviations from the measurement done by Yu's group due to the difference of electrolyte concentrations.[22]

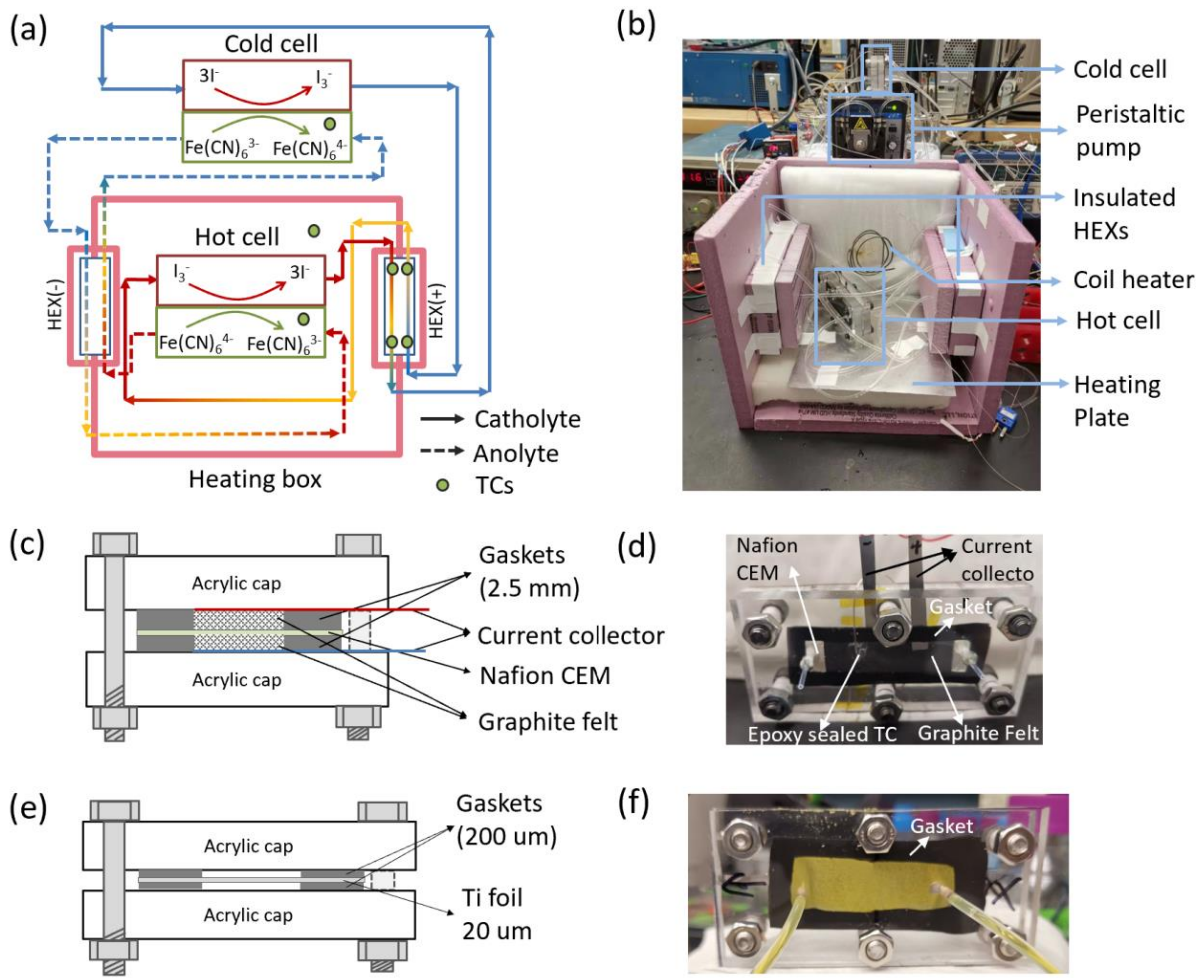

Figure 3. TRFB experimental testing system. (a) Schematic of the testing system, with temperature difference between the hot cell and cold cell achieved by a heating box chamber made with thermal insulation foams. (b) Picture of the dissembled heating box chamber showing the structure of temperature control system. (c) The schematic structure (not to scale) of the flow battery. Two acrylic caps were used as the structure host. The flow channel was created by silicone gaskets and filled with graphite felt as the porous electrode. A Nafion cation exchange membrane (CEM) was sandwiched between the gaskets and porous electrodes to separate the catholyte and anolyte. (d) The picture of the flow battery assembly. An epoxy sealed thermocouple as also inserted into the flow channel to monitor the temperature of the electrolyte. (e) The schematic structure (not to scale) and (f) the picture of planar heat exchanger.

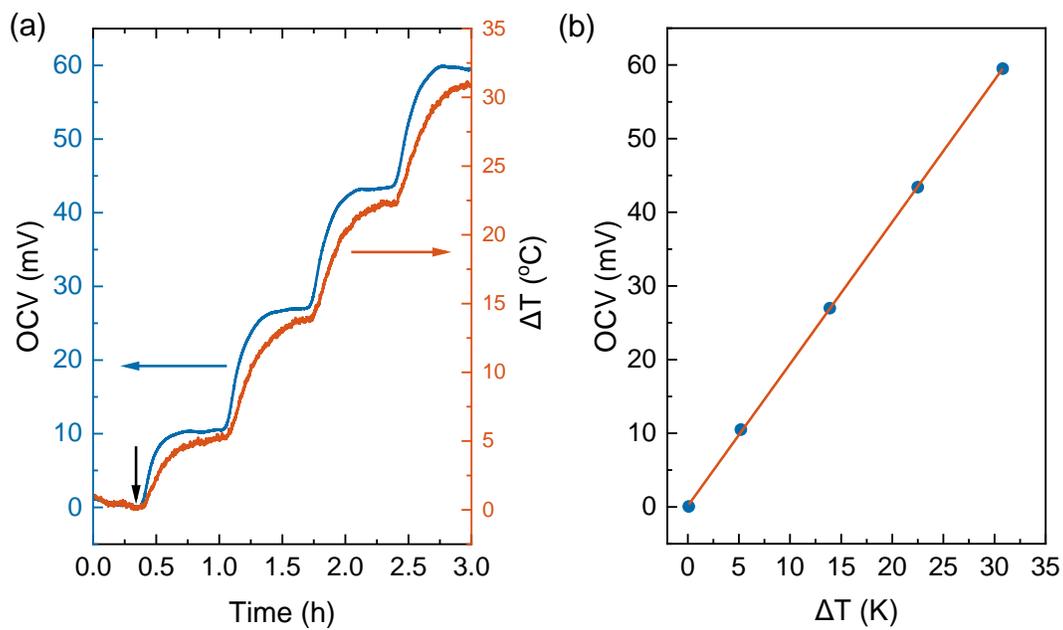

Figure 4. Temperature coefficient measurement of the TRFB using 0.1 M $I_2$ / 1 M KI as the catholyte, and 0.375M $K_3Fe(CN)_6$ / 0.375 M $K_4Fe(CN)_6$ as the anolyte. (a) Open circuit voltage (OCV) and temperature difference ($\Delta T$) as function of time. The vertical black arrow indicates the time start heating. (b) Temperature coefficient extracted by linear fitting the relation between OCV and $\Delta T$. The obtained temperature coefficient of the cell is $\alpha_{cell} = 1.9$ mV/K.

# 4. Results and Analysis of the TRFB's Performance

This section discusses the experimental results of flow rate dependent discharging performance of the TRFB cell, together with the theoretical analysis pointing to the factors limiting the efficiency and power density.

In Figures 5a-b, we first compared the measured flow rate dependent voltage and power density at $\Delta T = 37$ °C as a function of discharging current with the modeling results. Good agreement is achieved between the predicted and measured discharging behavior, indicating that our model accurately captures the coupled effect mass transfer, electrode polarization and reaction kinetics, which has been ignored in previous works. The parameters for the model prediction are summarized in Table 1. Only two parameters, the internal cell resistance $R_{in} = 7.25$ Ω and the tortuosity factor $\mathcal{T} = 5$ of the media, were fit from the data obtained at 230 μL/min (Re ≈ 3×10-3). This obtained tortuosity factor agreed reasonably with the literature values of graphite felts in the range of 5~6.[30] For other flow rates tested, we fixed these parameters and directly computed the results as plotted in Figure 5. In general, the mass transfer overpotential was larger at lower Reynolds numbers and flow rates, thus decreasing the power density. However, the extra amount of heat needed to increase the temperature of the electrolyte also decreased at smaller flow rate, and the efficiency $\eta_E$ increased at a smaller flow rate (Figure 5c). We estimated that 11% of $\eta_{Carnot}$ can be achieved at 1 μL/min at maximum power output, while the power density can approach 9.5 μW/cm2 at 0.5 mL/min. In general, simply choosing operating conditions (temperature difference $\Delta T$ and flow rate $Q$) of the TRFB cell cannot simultaneously optimize the power density and efficiency. Figure 6a-b show the map of peak power density and the corresponding efficiency as a function of temperature difference and flow rate. Although increasing temperature difference could simultaneously improve efficiency and power density, the effect of flow rate on power density and efficiency is opposite with nonperfect heat recuperation.

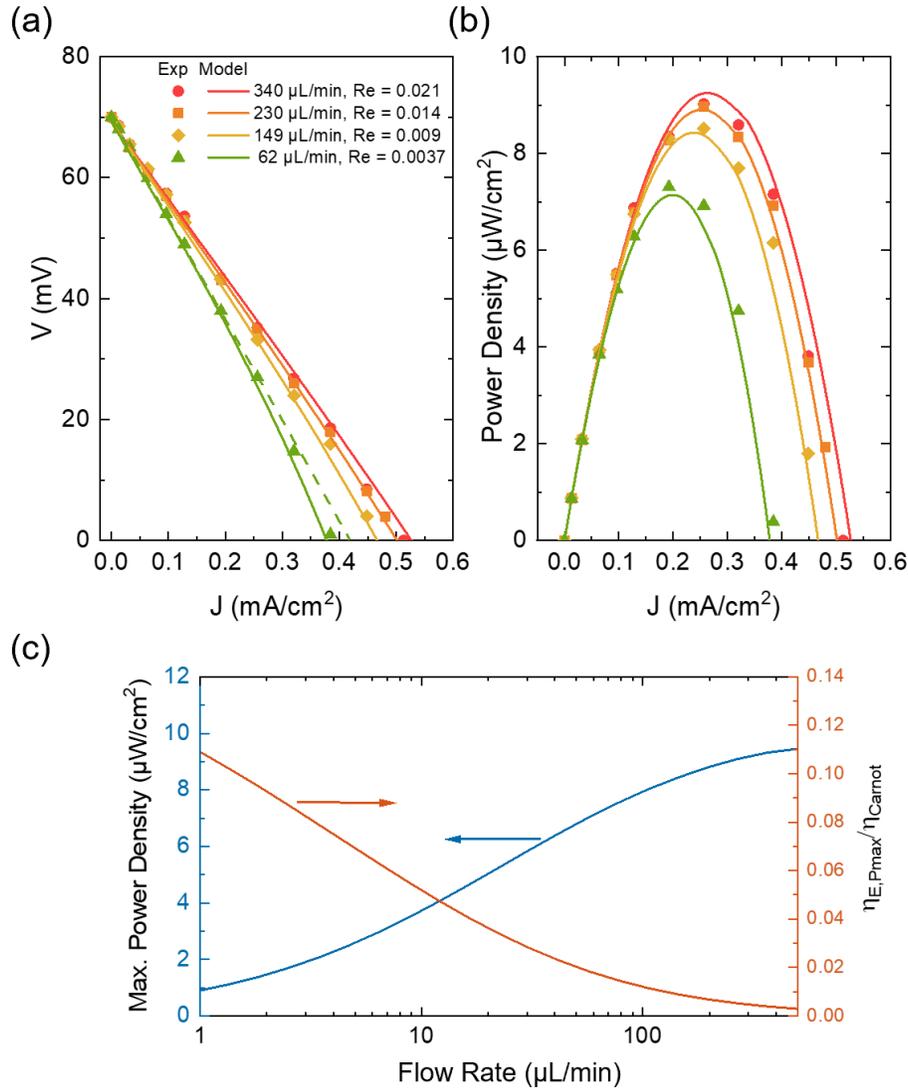

Figure 5. (a) Voltage and (b) areal power density as functions of discharge current $J$ at $\Delta T = 37°C$, with the hot cell kept at 60°C and the cold cell left at the ambient temperature (23°C). The symbols represent the experimental results and solid lines as modeling results. At a small flow rate (62 µL/min), current dependent discharging voltage deviated from the linear relation as indicated by the dashed green line. (c) Trade-off between the maximum power density and maximum relative efficiency $\eta_E/\eta_{Carnot}$ (in absolute scale from 0 to 1) as a function of flow rate.

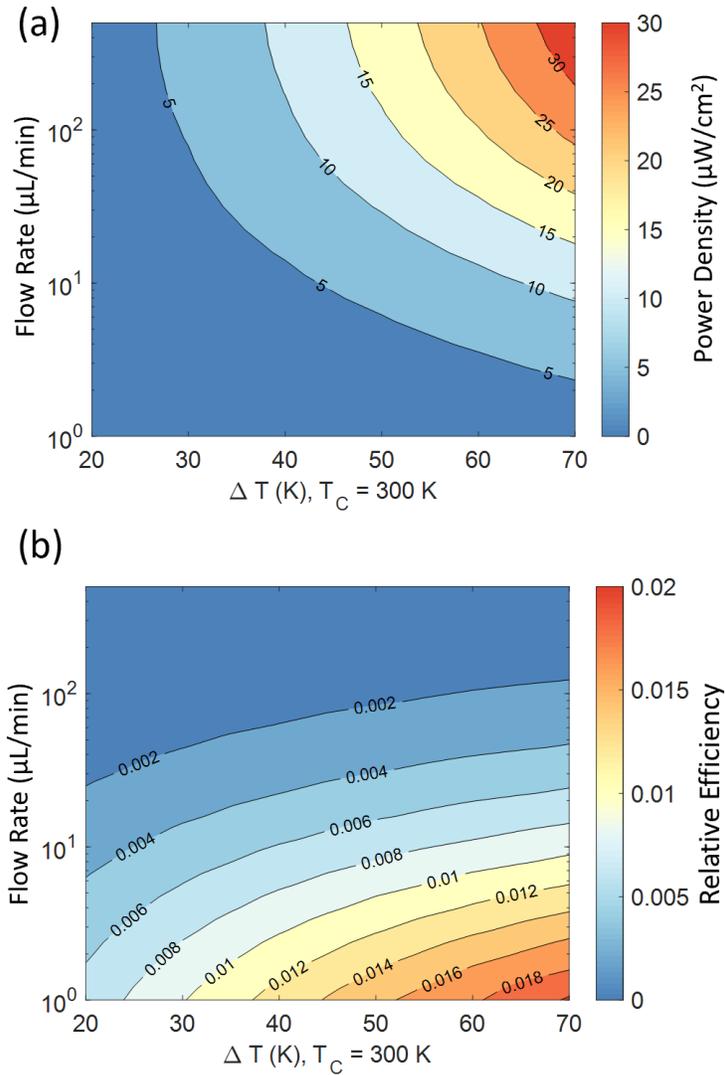

Figure 6. Map of (a) power density and (b) absolute efficiency (in absolute scale) as a function of temperature difference and flow rate, with heat recuperation effectiveness fixed at 90%.

Effectiveness of heat recuperation is pivotal to improving the energy efficiency of TRFB systems. Figure 7a-b show the map of absolute $\eta_{E,P_{max}}$ and relative efficiency $\eta_{E,P_{max}}/\eta_{Carnot}$ at the maximum power density, as a function of temperature difference $\Delta T$ and heat recuperation effectiveness $\epsilon_{HX}$. Figure 7b also highlights that the relative efficiency is very sensitive to the heat recuperation effectiveness, especially in the range $\epsilon_{HX} > 95\%$. With perfect recuperation, the relative efficiency could reach 50%, but decreases fast below 15% with $\epsilon_{HX}$ of 95%.

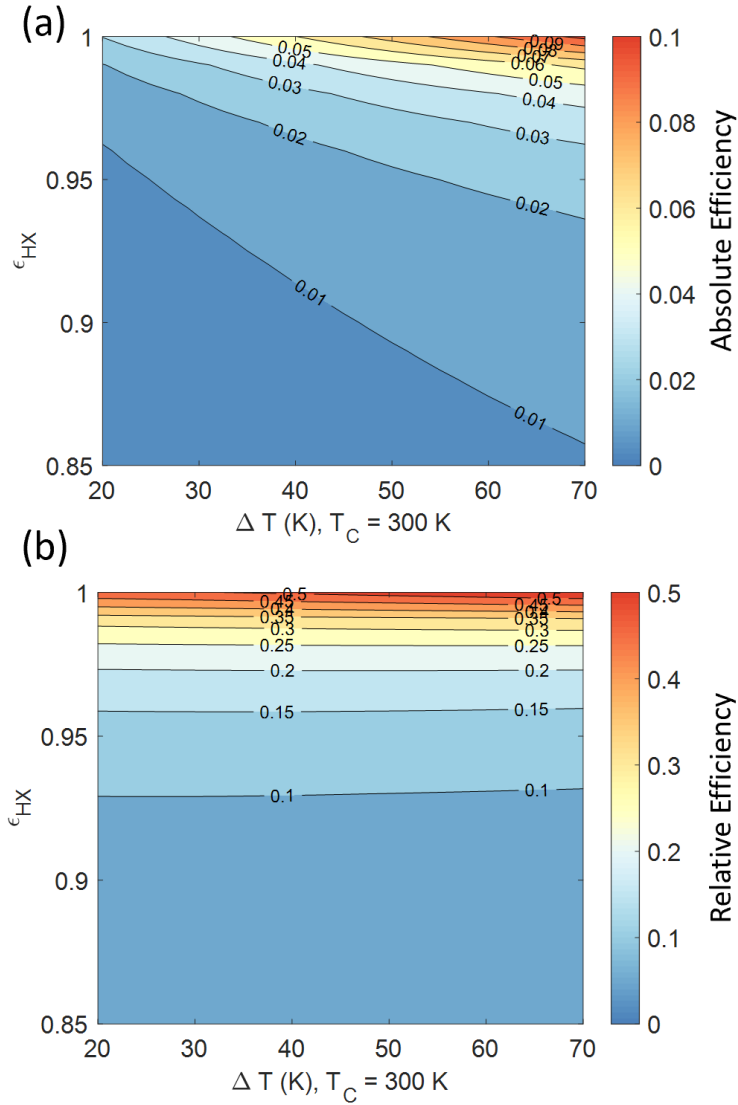

Figure 7. (a) Absolute and (b) relative efficiency map in absolute scale (from 0 to 1) at maximum power density as a function of heat recuperation effectiveness and temperature difference. Flow rate is fixed at 4 μL/min and the cold cell temperature is fixed at $T_C = 300$ K.

In addition to the operating parameters such as flow rate, temperature and heat recuperation, it is also important to understand the factors contributing to the overpotential as irreversible loss, which is important for designing TRFB cells to achieve higher efficiency and power density. Such irreversible loss is especially important when the flow rate is small, manifested in the clear deviation of the voltage drop from the linear voltage-current ($V - J$) relation at large discharge current (Figure 5a). Such nonlinear $V - J$ curve would result in lower thermodynamic efficiency. By pushing the effective ionic conductivity of the porous electrode $\sigma_{eff} \to \infty$, the nonlinearity of $V - J$ curve is eliminated. This loss due to the finite ionic conductivity inside the porous electrode is referred as the electrode polarization, as shown in Figure 8a. The major factor limiting both the power density and the efficiency is the mass transfer induced overpotential. By further pushing the Sherwood number to infinity, the relative efficiency can increase almost by two-fold from 7.5% to 14% (Figure 8b). Such analysis also shows that optimizing the electrochemical kinetics at the electrode might have negligible improvement on TRFB performances, if the other factors such as mass transfer becomes the dominant bottleneck.

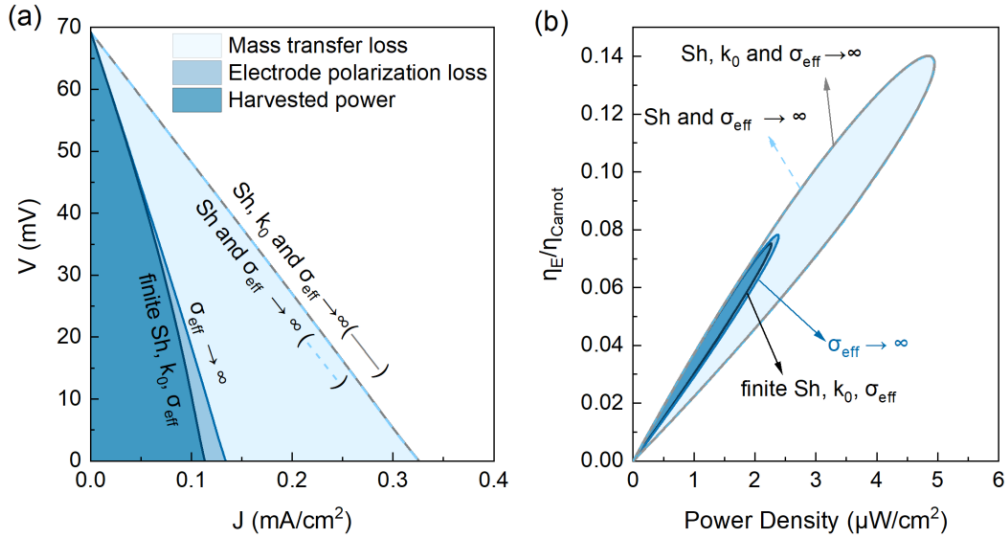

Figure 8. (a) Effect of the mass transfer overpotential loss, electrode polarization loss on the discharging voltage ($V$) – current ($J$) relation. The model is not sensitive to the electrochemical rate constant in the range of $k_0 > 0.75 \times 10^{-5}$ m/s. The discharging curve is almost unchanged by further letting the rate constant approaching $k_0 \to \infty$, based on discharge curve at the nonpolarized and infinite mass transfer limit ($Sh \to \infty, \sigma_{eff} \to \infty$). (b) The corresponding relative efficiency versus the power density. Efficiency is estimated assuming 90% heat recuperation effectiveness. This figure is estimated at the flow rate of 4 µL/min and the temperature difference of $\Delta T = 37$ °C.

Finally, the internal resistance ($R_{in}$) of the TRFB cells is also a major factor contributing to the loss, which is composed of the electrical ohmic resistance and the membrane resistances. Electrical resistance of the porous electrode and the contact resistance between the current collector and the electrode should not be the major contribution to $R_{in}$. We characterized the total electrical Ohmic resistance ($R_\Omega$) by measuring the resistance of the cell assembly without the Nafion membrane, showing that electrical Ohmic resistance is ~3 Ω. Since Nafion membrane has relative low ionic conductivity of K$^+$ (3.3×10$^{-4}$ S/cm),[37] the membrane resistance $R_{mem}$ was estimated to be ~ 5 Ω. Adding the $R_{mem}$ and $R_\Omega$ together, $R_{in}$ of a single cell was estimated to be around 8 Ω, agreeing with the $R_{in}$ obtained by fitting the experimental discharging curves in Table 1. The fitted $R_{in}$ value was slightly smaller, which might be attributed to the fact that electrodes were wetted by the electrolytes, and ionic conduction contributed to the extra conductance. If the membrane conductivity of $K^+$ can be improved close to that of proton $H^+$ ($7.8 \times 10^{-2}$ S/cm), then the membrane resistance would become negligible compared with the electrical Ohmic resistance. In this case, the power density at high flow rates (340 μL/min) can increase by two-fold from 9 μW/cm² to 18 μW/cm². However, the membrane resistance would have a smaller effect at low flow rates due to the limitation of the mass transfer. At the flow rate of 4 μL/min, the corresponding relative efficiency could be improved from 7.8% to 8.3%, while the power density increased by 14% at from 2.2 μW/cm² to 2.5 μW/cm².

Table 1. Summary of parameters used for the theoretical model

| Parameters | Values | Reference or Method |
|---|---|---|
| Internal resistance ($R_{in}$) | 7.25 Ω | Fitted to discharging curve |
| Porosity ($\epsilon$) | 95% | Ref. [38] |
| Tortuosity ($\mathcal{T}$) | 5.0 | Fitted to discharging curve |
| Fiber diameter ($d_f$) | 10 μm | Ref. [36] |
| Density of the catholyte ($\rho_+$) | 1191 kg/m$^3$ | Measured |
| Density of the anolyte ($\rho_-$) | 1282 kg/m$^3$ | Measured |
| Kinematic viscosity ($\nu$) | 8×10$^{-7}$ m$^2$/s | Ref. [39] |
| Diffusivity of KI$_3$ ($D_{O,+}$) | 7.0×10$^{-10}$ m$^2$/s | Ref. [22] |
| Diffusivity of KI ($D_{R,+}$) | 5.4×10$^{-9}$ m$^2$/s | Ref. [22] |
| Diffusivity of K$_3$Fe(CN)$_6$ ($D_{O,-}$) | 7.6×10$^{-10}$ m$^2$/s | Ref. [22] |
| Diffusivity of K$_4$Fe(CN)$_6$ ($D_{R,-}$) | 6.9×10$^{-10}$ m$^2$/s | Ref. [22] |
| Rate constant I$_3^-$/I$^-$ ($k_{0,+}$) | 6.7×10$^{-5}$ m/s | Ref. [22] |
| Rate constant of Fe(CN)$_6^{3-/4-}$ ($k_{0,-}$) | 0.75×10$^{-5}$ m/s | Ref. [22] |

## 5. Conclusions

In summary, we demonstrate a prototype of an all-anionic thermally regenerative flow battery operated with pH-neutral $K_3Fe(CN)_6/K_4Fe(CN)_6$ and $KI_3/KI$ electrolytes for the anolyte and catholyte solutions, respectively. A reasonably high temperature coefficient of the cell $\alpha_{Cell} = 1.9$ mV/K and a power density of 9 µW/cm$^2$ at $\Delta T = 37$ °C was measured. Such cell design is free of pH-matching issue affecting the stability of electrolytes and the issue of cross-over of active species since both active ions are anions. Combining the experiments with the porous electrode model, this work quantitatively captures the flow-rate dependent discharging behavior, even the nonlinear discharging characteristics at low flow rate or low Reynolds number. We showed that the efficiency relative to the Carnot limit can achieve 50% at the limit of perfect heat recuperation and 7.5% at 90% heat recuperation. For the first time, this work comprehensively analyzed the coupled effects of surface kinetics, mass transfer and electrode polarization on energy efficiency and power of TRFB for harvesting low-grade heat. At a fixed flow rate and internal resistance, mass transfer overpotential inside the porous electrode is the major factor limiting both the power density and efficiency. Electrode polarization becomes pronounced at low flow rates and large discharging currents, resulting in nonlinearity of the discharging voltage-current curves. Surface reaction kinetics characterized by the rate constant, however, does not significantly affect the performance of the system in the case when the mass transfer is the major limiting factor. We also expect that the power density and efficiency could be significantly improved in the future, with the development of proper cationic membranes with high conductivity or mobility of alkali metal cations to reduce the membrane resistance.

**Acknowledgment**

X.Q. acknowledges helpful and productive discussions with Dr. Haoran Jiang and Xun Wang. The authors gratefully acknowledge Prof. Evelyn Wang for valuable suggestions in the preparation of this manuscript. This work was supported by the Center for Mechanical Engineering Research and Education at MIT and SUSTech. Device modeling was partially performed using the Stampede2 supercomputer through the startup allocation (TG-DMR200043) of Extreme Science and Engineering Discovery Environment (XSEDE).  The authors declare no conflict of interests.